\documentclass[pre,aps,twocolumn,superscriptaddress,amssymb]{revtex4} %PRE format
\usepackage{epsfig,amsmath}
\usepackage{psfrag}

\begin{document}
\title{Time delay in the Kuramoto model with bimodal frequency distribution}

\author{Ernest Montbri\'o}
%\email{ernest.montbrio@upf.edu}
\affiliation{Computational Neuroscience, Technology Dept., Universitat Pompeu Fabra, 08003 Barcelona, Spain}
\affiliation{Departament de F\'{\i}sica, Universitat Illes Balears, 07122 Palma de Mallorca, Spain}
\affiliation{Institut Mediterrani d'Estudis Avan\c{c}ats IMEDEA (CSIC-UIB), 07122 Palma de Mallorca, Spain}
\author{Diego Paz\'o}
\altaffiliation{Present address: Instituto de F\'isica de Cantabria (CSIC-UC), Santander, Spain.}
\affiliation{Max-Planck-Institut f\"ur Physik komplexer Systeme, N\"othnitzer Stra{\ss}e 38, 01187 Dresden, Germany}
\author{J{\"u}rgen Schmidt}
\affiliation{Nichtlineare Dynamik, Institut f\"ur Physik, Universit\"at Potsdam, 14415 Potsdam, Germany}

\date{\today}

\begin{abstract}
We investigate the effects of a time-delayed all-to-all coupling scheme in a large population of oscillators with natural frequencies following a bimodal distribution. The regions of parameter space corresponding to synchronized and incoherent solutions are obtained both numerically and analytically for particular frequency distributions. In particular we find that bimodality introduces a new time scale that results in a quasiperiodic disposition of the regions of incoherence. 
\end{abstract}
\maketitle

The Kuramoto model~\cite{Kur84} is presumably the most successful attempt to study macroscopic synchronization phenomena arising in large heterogeneous populations of interacting self-oscillatory units~\cite{Str00,MMZ04,ABP+05}. Kuramoto, motivated by Winfree's work on biological oscillators~\cite{Win67}, showed that the dynamics of an ensemble of $N$ weakly interacting limit-cycle oscillators can be treated considering simply the oscillator phases $(\theta_1,\dots,\theta_N)$.
In this paper we study the Kuramoto model 
with delayed interactions~\cite{YS99}
\begin{equation}
\dot{\theta}_i(t)=  \omega_i - \frac{K}{N} \sum_{j=1}^{N} \sin[\theta_i(t)-\theta_j(t-\tau)] + \xi_i(t); \quad i=1,\dots, N,
\label{model0}
\end{equation}
where heterogeneity is established considering a certain distribution $g(\omega)$ of the natural frequencies $\omega_i$. The terms $\xi_i(t)$ represent uncorrelated zero-mean white noise processes, $\left< \xi_i(t)\right>=0$, $\left<\xi_i(t)\xi_j(t')\right>= 2D\delta_{ij}\delta(t-t')$. In absence of time delay ($\tau=0$) and for large $N$, 
as the coupling strength $K$ exceeds a critical threshold $K_c$ the model (\ref{model0}) shows drastically different transitions to collective synchronization, depending on the shape of $g(\omega)$~\cite{Kur84,Str00,MMZ04,Paz05,BNS92,Cra94}. 
For a strictly unimodal distribution the transition occurs between a totally incoherent state and a partially synchronized state. In contrast, symmetric bimodal distributions give rise to hysteresis and/or a transition to a time dependent state composed of two clusters~\cite{BNS92,Cra94}. 

The interactions in ensembles of coupled oscillators have been traditionally considered to be instantaneous, an assumption that considerably simplifies the analysis of such systems.  However, the study of phase oscillators with time-delayed coupling \footnote{A mathematically rigorous reduction from limit cycles with delayed interactions to phase oscillators is not fully justified~\cite{Izh98}, unless the time delay is considerably larger than the typical oscillation period. Nonetheless, important qualitative information is obtained considering both short and long delays altogether.} is receiving interest since a number of theoretical studies show that time delay may considerably affect the synchronization phenomena, typically leading to multistability of many synchronous states~(see e.g.~\cite{MMZ04,ABP+05} and references therein). In particular, the Kuramoto model with unimodal frequency distribution has been generalized to allow time-delayed interactions in \cite{YS99,CKK+00}. Additionally, phase models with time delay have successfully explained synchronization between plasmodial 
oscillators~\cite{TFE00} and in semiconductor laser arrays~\cite{KVM00}. 
Recent studies also demonstrate that time delay may be a useful 
synchronization-control mechanism in large oscillatory populations~\cite{RP04}.

In this paper we investigate the effects of a bimodal frequency distribution on the Kuramoto model (\ref{model0}) with time delay $\tau$. 
%%In absence of time delay 
For $\tau=0$ and assuming that $g(\omega)$ is symmetric
(centered at $\bar\omega$ with twin peaks of width $\gamma$ at both sides),
one may always transform to a rotating frame, 
such that the model~(\ref{model0}) is symmetric under the reflection: 
$(\theta_i,\omega_i) \rightarrow (-\theta_i,-\omega_i)$. 
Time delay generally breaks 
%%this 
such symmetry, 
except for the specific values $\tau=\tau_n \equiv n \pi/\bar\omega$ ($n \in \mathbb{N}$). 
%%Therefore, for these $\tau$ values
Thus, at $\tau=\tau_n$ we expect to recover the 
typical reflection-symmetric stationary wave solutions found in the Kuramoto model with bimodal frequency distribution but without delay~\cite{Cra94}. 
For general $\tau$ values the breaking of the reflection symmetry should give rise to the structures already observed for models without time delay but with either
asymmetric bimodal frequency distributions~\cite{ABD+98} or asymmetric coupling functions~\cite{MKB04}.

We begin our analysis by considering the noise-free case, $D=0$, and a frequency distribution that consists of two infinitely sharp peaks (i.e. $\gamma=0$):
\begin{equation} 
g(\omega)=[\delta(\omega-\omega_I)+\delta(\omega-\omega_{II})]/2
\label{g}
\end{equation}
with $\omega_{I}<\omega_{II}$. 
Even for this simple choice the results are illustrative and far from trivial.

\begin{figure}
\center
\includegraphics[width=3.4in,clip=true]{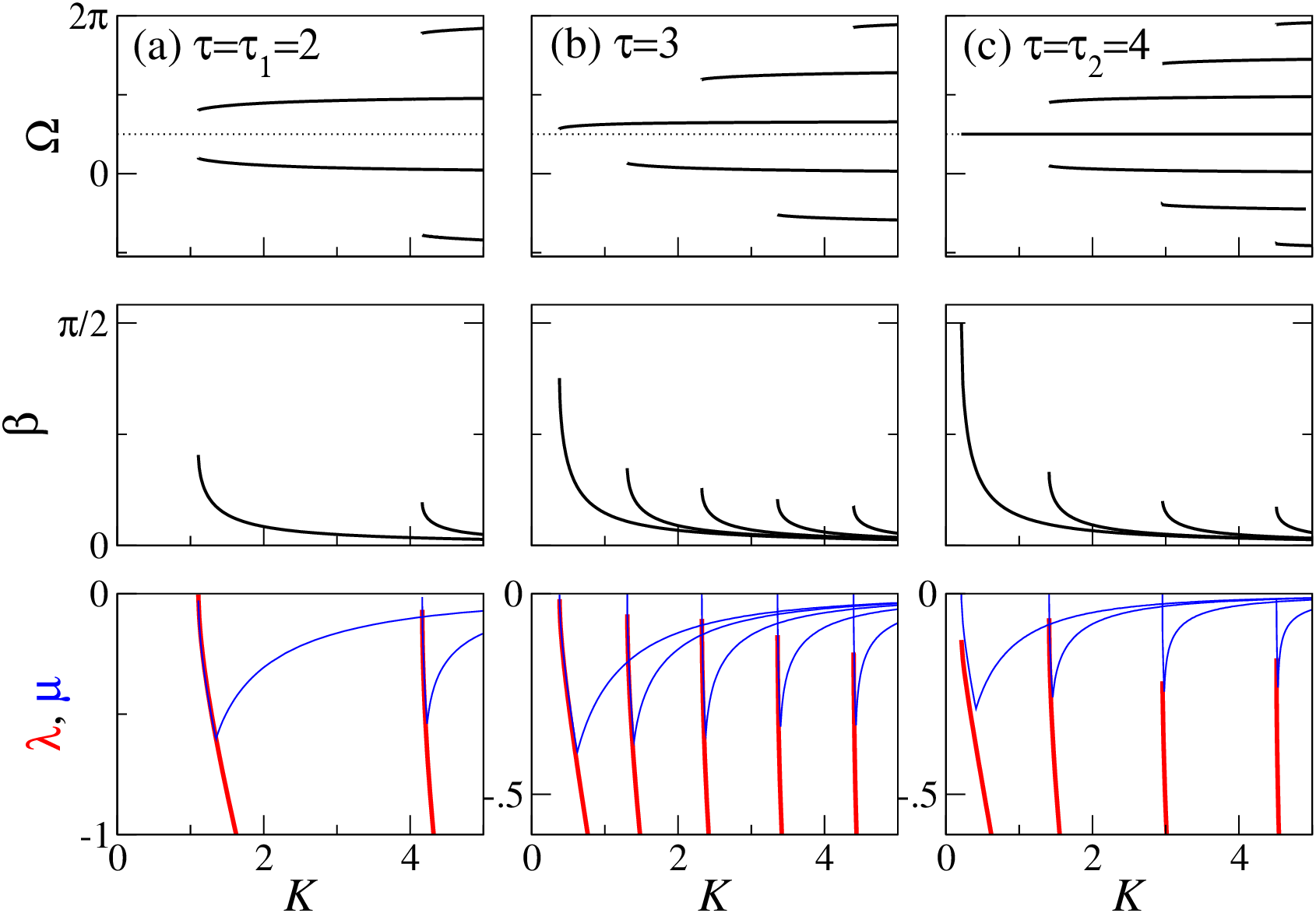}
\caption[]{(color online) Frequency $\Omega$, phase difference $\beta$ and leading Lyapunov exponents $\lambda$~(red), $\mu$~(blue), for the stable synchronized solutions (\ref{2sync}) as a function of $K$ for $\Delta \omega= \pi/15$, $\bar \omega=\pi/2$ (dotted line, upper panels).} 
\label{f1}
\end{figure}

\emph{Fully synchronized states.-} 
Let us first investigate the existence and stability of synchronized solutions 
of (\ref{model0}) consisting of two clusters of identical oscillators 
that rotate uniformly with angular velocity $\Omega$ and phase difference $\beta$
\begin{equation}
\Theta_{1,2} (t)= \Omega t \pm \beta/2,
\label{2sync}
\end{equation}
with $\Theta_1 \equiv \theta_{j\le N/2}$, $\Theta_2 \equiv \theta_{j>N/2}$ 
(taking $\omega_{j\le N/2}=\omega_I$, $\omega_{j>N/2}=\omega_{II}$). 
From ~(\ref{2sync}) and~(\ref{model0}), 
we find two transcendental equations for 
$\Omega$ and $\beta$ 
\begin{eqnarray}
\Omega &=& \bar\omega - K  \sin(\Omega\tau) [1+\cos\beta]/2, \label{cond1}\\ 
\sin\beta &=& \Delta \omega/ [K \cos(\Omega\tau)],   \label{cond2}
\end{eqnarray}
where $\Delta \omega \equiv \omega_{II}-\omega_{I}$ and $\bar\omega \equiv (\omega_I+\omega_{II})/2$. These equations have multiple solutions 
for given $K$, $\bar \omega$, $\Delta \omega$, and $\tau$ [see Fig.~(\ref{f1})].

The linear stability analysis of the  solutions~(\ref{2sync}) yields two $(\frac{N}{2}-1)$-fold degenerate eigenvalues
\begin{equation}
\lambda_\pm =-K [\cos(\Omega\tau) + \cos(\Omega\tau\pm \beta)]/2 ,
\label{eigenv1}
\end{equation}
and a set of eigenvalues $\{\mu_i\}$ which are determined by the transcendental equation
\begin{equation}
\begin{vmatrix} (1-e^{-\mu_i \tau}) c +c_{\scriptscriptstyle +}+\frac{2\mu_i}{K} &  -e^{-\mu_i \tau} c_{\scriptscriptstyle +} \\  -e^{-\mu_i \tau} c_{\scriptscriptstyle -} & (1-e^{-\mu_i \tau}) c + c_{\scriptscriptstyle -}+\frac{2\mu_i}{K} \end{vmatrix}=0, 
\label{eigenv2}
\end{equation}
where $c_{\scriptscriptstyle \pm} \equiv \cos(\Omega \tau \pm \beta)$, and $c \equiv \cos(\Omega \tau)$. The eigenvalues $\lambda_+,\lambda_-$ govern the stability {\it within} the two individual clusters, each with $\frac{N}{2}$ oscillators. The eigenvalues $\mu_i$ (discarding the trivial solution $\mu_0=0$) 
are related to the stability of the frequency locking {\it between} the two clusters. 
% The stability of these inter-cluster modes is determined via Eq.~(\ref{eigenv2}) yielding an infinite number of relaxation modes 
% (as expected, since the system has infinite dimensions due to time delay). 
% By contrast, the modes of stabilization of each cluster are governed by the simpler algebraic equations~(\ref{eigenv1}). 
We denote the leading 
Lyapunov exponents in each subset 
by $\lambda\equiv\max\{\lambda_+,\lambda_-\}$ and $\mu\equiv\max\{\text{Re}(\mu_i)\}~(i \neq 0)$. 
 
Figure~\ref{f1} displays numerical solutions of Eqs.~(\ref{cond1}), (\ref{cond2}),
(\ref{eigenv1}) and (\ref{eigenv2}) 
for specific values of the parameters $\tau$, $\Delta \omega$ and $\bar\omega$. 
The upper panels show that for $\tau=\tau_n \equiv n \pi/\bar\omega$ symmetry-related solutions with frequencies $\Omega(K,\tau_n)=\bar\omega \pm \zeta(K,\tau_n)$ arise in pairs around the ``central'' frequency $\bar{\omega}$ (which is the synchronization frequency for $\tau=0$) [Fig.~\ref{f1}~(a,c)]. For positive $K$, the central solution is stable (unstable) for even (odd) values of $n$, and vice versa for negative $K$. For intermediate values, $\tau_n <\tau < \tau_{n+1}$, the stable solutions continuously vary between these two patterns, all of them approaching $\Omega=0$ as $\tau$ increases. This effect, common to time-delayed interacting oscillatory systems, is known as frequency suppression~\cite{NSK91}.

The growing number of synchronized solutions as $K$ and $\tau$ are increased 
(see Fig.~\ref{f1}) was already reported by Schuster and Wagner~\cite{SW89} for a system of two coupled oscillators, corresponding to model~(\ref{model0}) for $N=2$ without the self-coupling term.
% ~\footnote{Coupling within a cluster arises naturally for a 
% population of oscillators. The overall effect of a single oscillator coupled to itself is negligible in the limit $N\rightarrow\infty$. 
%%The self-coupling arises naturally in a population except for each single oscillator with itself, what is negligible in the $N\rightarrow \infty$ limit.
%}. 
They also have shown that as $K$ is increased, stable solutions alternately appear with small ($\beta\approx0$) and large ($\beta\approx\pi$) phase differences;
this is also the case in the presence of self-coupling~\footnote{B.~Lysyansky (private communication).}. However, for a population---due to the existence of 
the 
%instabilization 
destabilizing
modes linked to~(\ref{eigenv1})---all the stable solutions 
are of the in-phase type, i.e., $0<\beta<\pi/2$~[see Fig.~\ref{f1}, central panels].
These solutions appear at saddle-node bifurcations where either $\lambda$ or $\mu$ vanishes~[see Fig.~\ref{f1}, lower panels]. 
Increasing $K$, the Lyapunov exponent $\lambda$ is stabilized ($\lambda \propto -K$) whereas $\mu$ first decreases steeply and then abruptly changes its tendency 
(at the point where a pair of complex conjugate eigenvalues $\mu_i$ become the leading ones), and asymptotically approaches zero. 

\begin{figure}
\center
\includegraphics[width=3.2in,clip=true]{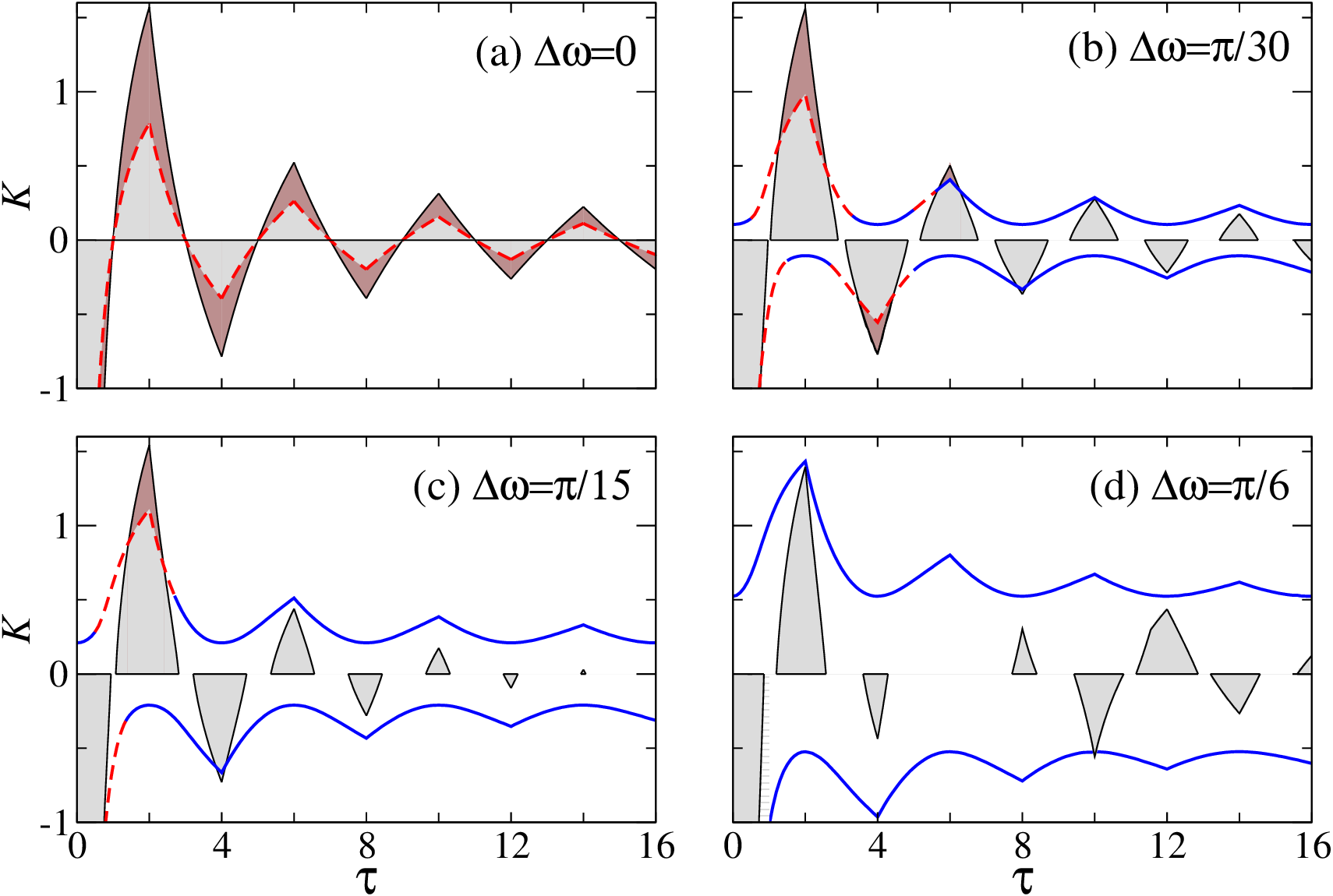}
\caption[]{(color online) Stability boundaries for \emph{Synchronization} and \emph{Incoherence}.
Theoretical boundaries for (full) synchronization are obtained from Eqs.(\ref{eigenv1},\ref{eigenv2}). Disintegration ($\lambda \rightarrow 0^-$) and  unlocking ($\mu \rightarrow0^-$) are indicated with red-dashed  and blue-solid lines, respectively.  
Black-thin curves: Boundaries of stable incoherence [Eq.~(\ref{kcpm})]. Shaded areas: Incoherence regions obtained by numerical integration\footnote{Using a 
fifth order Adams-Bashforth-Moulton scheme with $dt=\tau/20$, except for $\tau<1$, where $dt=0.05$ , $N=24$, $\bar \omega=\pi/2$.} of Eqs.~(\ref{model0}). Dark-shaded regions: bistability between incoherence and synchronization.
}
\label{diagr}
\end{figure}

Figure~\ref{diagr} shows the boundaries of the synchronous states (\ref{2sync}) in the $\tau-K$ plane.
For a unimodal distribution ($\Delta\omega=0$) the regions without stable synchronous solutions are
disconnected from each other, and the instability
is always via the disintegration of the cluster
[red-dashed lines, Fig.~\ref{diagr}(a)]. By contrast, for a bimodal distribution ($\Delta\omega>0$) the stability boundaries detach from the $K=0$ axis and we obtain two separated continuous curves of marginal stability for $K<0$ and $K>0$. Note that as $\Delta \omega$ is increased the instability 
of the phase locked state occurs 
mostly via unlocking of the two clusters.

\emph{Incoherent state.-} 
Next we address an important type of solution of Eqs.~(\ref{model0}), 
so-called incoherence, in which the system is in a completely phase-disordered state. To study the stability of the incoherent state we introduce the 
complex order parameter $R e^{i\psi}= N^{-1} \sum_{j=1}^{N}e^{i\theta_j}$, that measures 
the degree of ``phase coherence" in the system. This permits to write the system~(\ref{model0}) 
in terms of the time-delayed mean field quantities $R$ and $\psi$~\cite{Kur84}    
\begin{equation}
\dot\theta_i(t) = \omega_i  - K R(t-\tau) \sin[\theta_i(t)-\psi(t-\tau)] + \xi_i(t).
\label{model}
\end{equation}
Considering the limit $N\rightarrow \infty$, we drop the indices 
and introduce the probability density for the phases $\rho(\theta,t,\omega)$~\cite{SM91}. Then $\rho$ obeys the Fokker-Planck equation $\partial \rho/\partial t= - \partial(\rho v)/\partial \theta +D \partial^2\rho /\partial \theta^2 $ (where $v=\omega - K R(t-\tau) \sin[\theta(t)-\psi(t-\tau)]$), for which  the incoherent state $\rho_0=(2\pi)^{-1}$ is always a trivial stationary solution. Linearizing the Fokker-Planck equation about $\rho_0$ one finds that the stability of incoherence is determined by the eigenvalues $\Lambda$ satisfying~\cite{YS99} 
\begin{equation}
e^{-\Lambda \tau} \frac{K}{2}\int_{-\infty}^{\infty} \frac{g(\omega)}{\Lambda+D+i\omega} d\omega=1.
\label{incoherence}
\end{equation}
This equation with the distribution (\ref{g}) 
and $D=0$ gives (after setting $\Lambda=-i\Omega_c$, to obtain the instability threshold)  
the collection of critical curves
\begin{eqnarray}
K^{(l)}_c &=& (-2)^l (\omega_I-\Omega^{(l)}_c) (\omega_{II}-\Omega^{(l)}_c) /(\bar\omega-\Omega^{(l)}_c), \label{kc} \\
\Omega^{(l)}_c&=&(1/2+l)\pi/\tau,
\end{eqnarray}
($l$ integer), and two trivial ones $K_c=0$ (for $\Omega_c=\omega_I$ and $\omega_{II}$). This leads us to the bounds for stable incoherence 
\begin{subequations}
\label{kcpm}
\begin{eqnarray}
K^{+}_c(\tau) &=& \min\{K^{(l)}_c(\tau)\big| K^{(l)}_c(\tau)>0\} , \\
K^{-}_c(\tau) &=& \max\{K^{(l)}_c(\tau)\big| K^{(l)}_c(\tau)<0\} .
\end{eqnarray}
\end{subequations}
Figure~\ref{diagr}~(a) shows the regions of incoherence obtained from Eq.~(\ref{kcpm}) for a unimodal distribution ($\Delta\omega=0$)~\cite{YS99}. They consist of a periodic disposition of disconnected tent-shaped regions centered at $\tau=\tau_n$, of height $K_c=(-1)^{n+1} \pi/\tau_n$ and width $\tau_1$ (at $K=0$), as can be seen by setting $\bar\omega=\omega_I=\omega_{II}$ in Eq.~(\ref{kc}). 

In contrast, the regions of incoherence for  
the bimodal frequency distribution (\ref{g}) 
are organized quasiperiodically, with a decaying envelope of period $2\pi/\Delta\omega$ ~[see Fig.~\ref{diagr}(b-d)]. 
This can be understood noticing that, for $|K| \approx 0$, such regions correspond to the overlapping of the incoherence regions of 
two independent unimodal populations (with $\bar \omega=\omega_I$ and $\omega_{II}$). 
Thus, the new disposition of the regions of incoherence is quasiperiodic 
with frequencies $\omega_I$ and $\omega_{II}$, and it can be determined through a function  %
\begin{equation}
f(\tau)=\cos(\omega_I \tau)\cos(\omega_{II} \tau) \propto \cos(2 \bar\omega  \tau)+\cos(\Delta\omega \tau),
\label{f}
\end{equation}
which takes positive values exactly where incoherence is stable for some $K \neq 0$. 

\begin{figure}
\center
\includegraphics[width=3.in,clip=true]{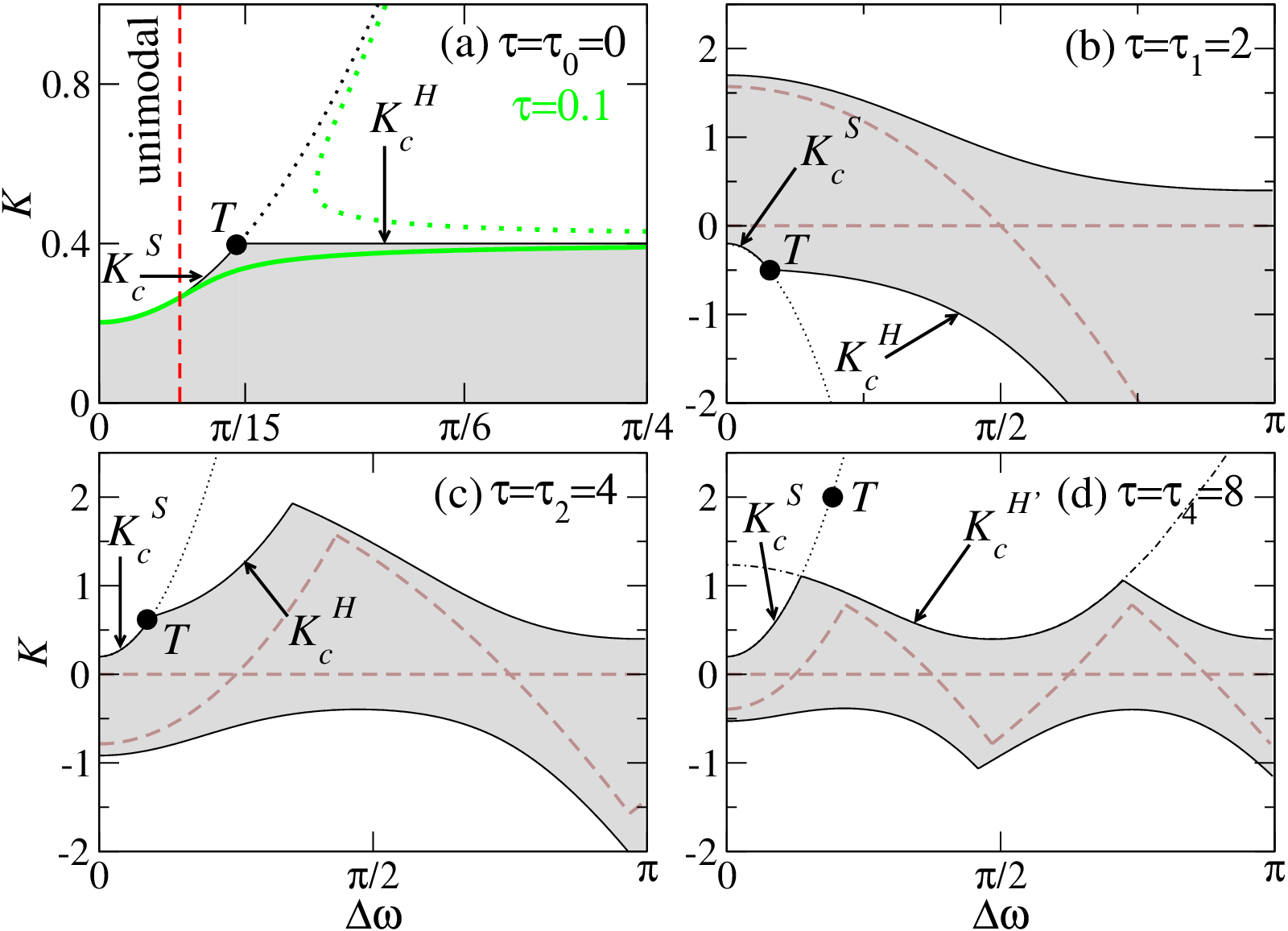}
\caption[]{(color online) $\Delta\omega-K$ diagram 
showing the regions of stable incoherence for a bi-Lorentzian frequency 
distribution of width $\gamma=0.1$ centered at $\bar \omega=\pi/2$. Solid-black ($\gamma=0.1$) and brown-dashed ($\gamma=0$) curves: analytical boundaries obtained from Eqs.~(\ref{eigenv_i}) and (\ref{kcpm}), respectively. Shaded areas:  Regions of incoherence obtained by numerical integration of Eqs.~(\ref{model0}), $\gamma=0.1,~D=0,~N=2000$. (a) Green curves ($\tau=0.1$): symmetry breaking which results in two Hopf bifurcations (the solid green curve becomes the new boundary). To the right of the red-dashed line at $\Delta\omega=2\gamma/\sqrt{3}$  [only shown in panel (a)] the bi-Lorentzian distribution is bimodal.}
\label{f3}
\end{figure}

\emph{Effect of frequency diversity (and noise).-}
For a unimodal frequency distribution the inclusion of diversity (or noise) does not alter 
significantly the scenario already captured for identical oscillators~\cite{YS99}, but for the bimodal distribution it has a more intricate effect. 

We restrict our analysis to the bi-Lorentzian distribution,
but a qualitatively similar scenario is expected for other bimodal 
distributions.
Specifically, we take  $g(\omega)=\gamma/(2\pi)( [(\omega-\omega_I)^2 + \gamma^2]^{-1}+[(\omega-\omega_{II})^2 + \gamma^2]^{-1})$ 
which, for $\Delta\omega> 2 \gamma /\sqrt{3}$, becomes bimodal. Then the linear stability of the incoherent state, determined via Eq.~(\ref{incoherence}), yields the transcendental equation 
\begin{equation}
\Lambda_\pm=-D-\gamma+\frac{1}{4}\left[K e^{-  \Lambda_\pm \tau}\pm \sqrt{K^2 e^{- 2 \Lambda_\pm \tau}  -4 \Delta\omega^2} \right]-i \bar\omega.
\label{eigenv_i}
\end{equation}
$D$ and $\gamma$  appear at linear order always in the combination $(D+\gamma)$, 
hence both effects can be studied together restricting to the case $D=0$. 

With the help of Fig.~\ref{f3}(a), let us first recall the main results for $\tau=0$~\cite{Cra94}. For small separation of the peaks ($2\gamma/\sqrt{3}<\Delta\omega< 2\gamma$), incoherence becomes unstable \emph{subcritically} in a steady-state bifurcation at $K_c^S=2[(\Delta\omega/2)^2+\gamma^2]/\gamma$
\footnote{This is the classical result of Kuramoto: $K_c=2/\pi g(\bar\omega)$, 
with subcriticality due to $g''(\bar\omega)>0$~\cite{Kur84}.}. 
In this case the system exhibits a single cluster of oscillators locked to the frequency $\Omega_c=\bar\omega$. If $\Delta\omega> 2\gamma$, a degenerate Hopf bifurcation at $K_c^H=4\gamma$ precedes the previous instability. 
At $K_c^H$ two symmetric clusters at both sides of the central frequency $\bar\omega$ appear simultaneously. 
The loci of both bifurcations coalesce at $T$ (a double-zero eigenvalue point in the co-rotating frame).

A similar scenario is recovered periodically at the values $\tau=\tau_n$, 
i.e., when system~(\ref{model0}) has reflection symmetry. The steady-state bifurcation occurs at the same values of $K$, but with a change of sign for odd $n$
\begin{equation}
K_c^S(\tau=\tau_n)=(-1)^n ~2[(\Delta\omega/2)^2+\gamma^2]/\gamma.
\label{kct0}
\end{equation}
Beyond the codimension-two point $T$, located on the curve $K_c^S$,
at $\Delta\omega=2\gamma\sqrt{(1+\gamma\tau_n)/(1-\gamma\tau_n)}$, this instability 
is preceded by a degenerate Hopf bifurcation [see Fig.~\ref{f3}(b,c)]. 
The latter expression diverges at $\tau=\gamma^{-1}$ and hence, above a certain value of $n$ ($n\ge 4$ for the parameters in Fig.~\ref{f3}), another degenerate Hopf bifurcation, denoted by $K_c^{H'}$ in Fig.~\ref{f3}(d) (dash-dotted line), crosses the $K_c^S$ line and prevents incoherence to be stable for larger $K$ \footnote{The bifurcation $K_c^{H'}$ is not contained in the center manifold of $T$ ($\Omega_c \nrightarrow \bar\omega$  as $K \rightarrow K_c^S$ along $K_c^{H'}$).}.
\begin{figure}
\center
\includegraphics[width=3.2in,clip=true]{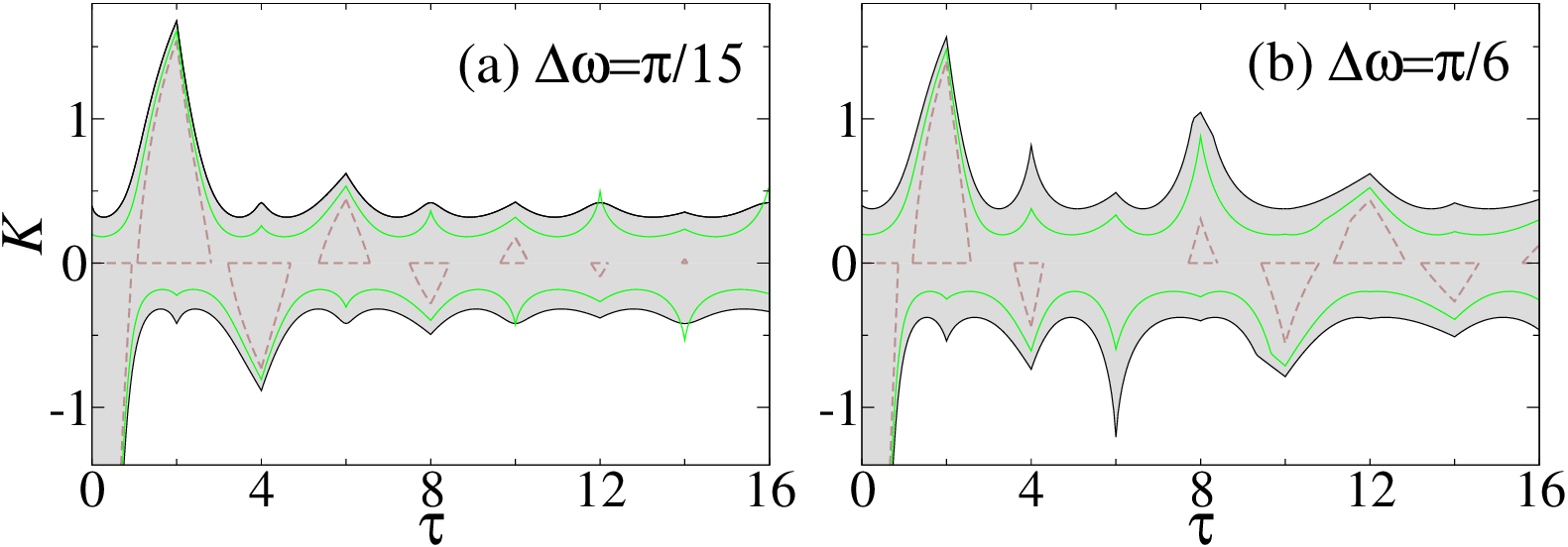}
\caption[]{(color online) $\tau-K$ diagram with the stable regions of the incoherent
 state for the bi-Lorentzian frequency distribution for $\gamma=0$ [dashed brown, 
from Eq.~(\ref{eigenv_i})], and [from Eq.~(\ref{kcpm})] for $\gamma=0.05$ (solid
green) and $\gamma=0.1$ (solid black). Shaded area: Incoherence regions obtained 
by numerical integration of Eq.~(\ref{model0}), ($\gamma=0.1,D=0,N=2000$).}
\label{f4}
\end{figure}

A general value 
%for 
of
the time delay, $\tau\neq\tau_n$, changes this scenario 
due to the breakdown of reflection symmetry. For $\tau$ close to $\tau_n$, the
Hopf bifurcations continuously split, as it is shown in Fig.~\ref{f3}(a) for $\tau=0.1$. This occurs in a similar way as in previous works studying the breakdown of reflection symmetry without time delay~\cite{MKB04}. Generally, such symmetry breaking implies asymmetric non-simultaneous nucleations, 
as well as a complete modification of the bistability types and transitions to synchronization.

Figure~\ref{f4} shows the region of incoherence in the $\tau-K$ plane for two 
values of $\Delta\omega$. Due to the splitting of the degenerate Hopf bifurcations
as $\tau$ is shifted from $\tau_n$, local extrema of incoherence appear exactly at $\tau=\tau_n$. Actually, the overall picture is more complex for arbitrary values of $\gamma$, since some peaks are washed out, and different bimodal distributions may show specific shapes. However, in the $\gamma \rightarrow 0$ limit, we provide a general distribution-independent statement (suggested from numerical solutions of Eq.~(\ref{eigenv_i}) and analytical arguments): peaks will appear at all $\tau=\tau_n$ values, unless there is a resonance ($\Delta\omega/\bar\omega = p/q$) what
excludes $n= 2 m q/p$, $m$ even (odd) for positive (negative) $K$ values. 

\emph{Conclusions.-} We have studied the Kuramoto model with time delay by analyzing the linear stability properties of the synchronized and incoherent 
solutions for bi-delta distributions~(\ref{g}).
Our results have been compared to previous studies on phase oscillators 
coupled via time-delayed interactions. In particular, in contrast to the $N=2$ case~\cite{SW89}, the synchronous states only admit small phase differences.
Also, compared to unimodal distributions~\cite{YS99}
stability boundaries of the incoherent state have 
a completely different structure in parameter space, due to the presence of a 
new time scale in the system which results in a quasiperiodic pattern. 
Finally, we have considered the effect of diversity on the incoherent state exploiting the reflection symmetry of the Kuramoto model model for certain values of the time delay. For such values, we have achieved some analytical results 
that extend the non-delayed case studied in~\cite{Cra94}, and that  
help to build up a general picture for the shape of the regions of incoherence. These results may be of interest for the study of dynamical networks with bimodal frequency distribution, where the transmission times may play an important role on the synchronization dynamics, e.g.~in neuroscience. For example, in the visual cortex, different populations of neurons form  spatially separated cortical columns that interact with a significant 
time delay. In this context, a successful phase-model for visual processing was 
proposed in~\cite{SGK91}, but the time delay effects were only studied for two oscillators.   

We thank Claudio Tessone, Bernd Blasius and J\"urgen Kurths for fruitful discussions. E.M. was partially supported by the European research project EmCAP (FP6-IST, contract 013123).

\end{document}